\documentclass[aps,prl,superscriptaddress,preprint]{revtex4-1}
\usepackage{graphicx}
\usepackage{verbatim}
\usepackage{amsmath,amssymb}
\begin{document}
\newcommand{\orf}{$\Omega_{\textup{rf}}$}
\newcommand{\op}{$\Omega_{\textup{p}}$}
\newcommand{\omc}{$\Omega_{\textup{c}}$}
\newcommand{\ket}[1]{$\vert #1 \rangle$}
\newcommand{\figwidth}{3in}

\title{Quantum Assisted Electrometry using Bright Atomic Resonances}

\author{J. Sedlacek}
\affiliation{Homer L. Dodge Department of Physics and Astronomy, The University of Oklahoma, 440 W. Brooks St. Norman, OK 73019, USA}
\author{A. Schwettmann}
\affiliation{Homer L. Dodge Department of Physics and Astronomy, The University of Oklahoma, 440 W. Brooks St. Norman, OK 73019, USA}
\author{H. K\"{u}bler}
\affiliation{Homer L. Dodge Department of Physics and Astronomy, The University of Oklahoma, 440 W. Brooks St. Norman, OK 73019, USA}
\affiliation{5. Physikalisches Institut, Universit\"{a}t Stuttgart, Pfaffenwaldring 57 D-70550 Stuttgart, Germany}
\author{R. L\"{o}w}
\affiliation{5. Physikalisches Institut, Universit\"{a}t Stuttgart, Pfaffenwaldring 57 D-70550 Stuttgart, Germany}
\author{T. Pfau}
\affiliation{5. Physikalisches Institut, Universit\"{a}t Stuttgart, Pfaffenwaldring 57 D-70550 Stuttgart, Germany}
\author{J.P. Shaffer}
\email[]{shaffer@nhn.ou.edu}
\affiliation{Homer L. Dodge Department of Physics and Astronomy, The University of Oklahoma, 440 W. Brooks St. Norman, OK 73019, USA}

\begin{abstract}
We demonstrate a new method for measuring radio frequency (RF) electric fields based on quantum interference in an atom. Using a bright resonance prepared within an electromagnetically induced transparency window we are able to achieve a sensitivity of $\sim 30\,\mu$V$\,$cm$^{-1}\,\sqrt{\mathrm{Hz}}^{-1}$. For this work, we demonstrated detection of RF electric fields as small as $\sim 8\,\mu$V$\,$cm$^{-1}$. The sensitivity is currently limited by the spectral bandwidth of the lasers used in the experiment and can be significantly improved in the future. The method can serve as a new quantum based standard for RF electrometry. The reproducibility, accuracy and stability of using a quantum system for measuring RF electric fields promises to advance electrometry to the current levels of magnetometry.
\end{abstract}

\maketitle

The pursuit of quantum based standards for the units of length and time as well as measurements of interesting physical quantities, such as the magnetic field, has been a major research direction for some time because quantum systems show clear advantages for providing stable and uniform measurements. The improvement of standards is linked to the accurate measurement of fundamental physical constants such as $\hbar$, $c$ and the properties of fundamental particles, which are also improved by using quantum systems. The reproducibility, accuracy and stability of the quantum world has made many of the standards and measurement techniques established early in the last century relics,  notable exceptions being mass and electric field. The most progress in adopting quantum systems for standards has been made in measuring atomic transition frequencies that are now used for length and time standards. Atomic clocks have now realized an accuracy of better than 1 part in $10^{14}$ \cite{clockreview}. Concurrently, tremendous progress has also been realized in measuring magnetic fields, where it is now possible to achieve an accuracy of fT$\,\sqrt{\mathrm{Hz}}^{-1}$ \cite{SERF,Budker,Polzik,mag}. Motivated by applications of these advances, such as g factor measurements \cite{Gabrielse} and the global positioning system, it is straightforward to conclude that quantum assisted sensing is relevant both fundamentally and technologically. It is interesting and meaningful to explore the possibility of extending the idea of quantum assisted sensing to other important physical quantities.

In contrast to the successes already mentioned, comparatively little progress has been made in applying quantum systems to measure the electric field, despite the clear advantages of using them for standards and the ubiquity of the electric field. The basic concepts used to measure a radio frequency (RF) electric field and the standards for calibrating those devices have changed little from the ones Hertz pioneered in the 1880's \cite{Russianreview}. Hertz implemented dipole antennas to experimentally establish the existence of electro-magnetic waves. Standards from the last century focussed on specifying antenna geometry and placement in addition to developing the theory of these devices. The current standards for electric field measurement are referred to as the 'Standard Antenna' and 'Standard Field' methods \cite{NBSstandards,NIST1}. For fields up to $40\,$GHz, these techniques use a resistively loaded dipole antenna and microwave diode detector. Electric fields at a level of $\sim 1\,$mV$\,$cm$^{-1}$ can be determined and used for calibration \cite{NIST2,Russianstandards,Russian2,Russian3}. Modern variations on sensing electric fields for traceable standards are based on optical measurements of the electro-magnetic fields converted by an antenna. In particular, the Pockels effect can be used. These setups can sense electric field strengths down to $\sim 30\,\mu$V$\,$cm$^{-1}$ \cite{Koysemi}. In all these types of measurements, the voltage induced in the probe has to be used to back calculate the electric field. For electric field sensing, then, a major limitation is the antenna as it is the converter of the electric field to some observable, for example voltage, and depends on geometry, can lead to perturbations of the electric field, particularly in the near field, and can suffer from out of band interference.

The accurate measurement of electric fields is important for a host of applications, especially at RF frequencies. Perhaps most importantly, the ability to sensitively measure RF electric fields can allow for the amplitude stabilization of an RF electric field source and the determination of optical properties of materials at these frequencies to high precision. Prior work in this direction has been termed the development of an 'atomic candle' \cite{atomiccandle,atomiccandle2}.

Here, we demonstrate a method for making a traceable electric field sensor based on Rydberg atoms excited in a glass vapor cell. The working principle of the sensor is based on detecting how RF electric fields affect the optical transitions of Rb Rydberg atoms. The atoms are each setup as a quantum interferometer using electromagnetically induced transparency (EIT) and the RF electric field is detected as a bright resonance within the EIT signal. We demonstrate a sensitivity of $\sim 30\,\mu$V$\,$cm$^{-1} \sqrt{\mathrm{Hz}}^{-1}$ which is currently limited by the laser spectral bandwidths used for the experiments. We show that RF electric field amplitudes as low as $\sim 8\,\mu$V$\,$cm$^{-1}$ can be observed. The minimum detected RF electric field is already superior to current methods. Optical readout allows for spatial resolution in the micrometer regime and facilitates near field measurments. The basic concept can be scaled up to larger one or two dimensional arrays that include fiber optical delivery of the light for readout. The setup and detection scheme is conducive for miniaturization, particularly using microcells \cite{Kubler10}. The technique presented here can also yield information about the polarization of the field but the details of this aspect of the method lie outside the scope of this work. We will discuss the polarization dependence in a future publication.

 \section{Approach}

EIT is a quantum interference process where two excitation pathways of a 3-level system interfere to produce transmission on an atomic resonance \cite{fleisch05}. The quantum interference produces a dark state that does not absorb light under normal resonance conditions. Since EIT depends on quantum interference between 2 excitation pathways, EIT is exquisitely sensitive to phase disturbances, transitions out of the participating states and energy level shifts of the 3 level system. The addition of an RF electric field that is resonant with a nearby Rydberg transition, a 4th level, for the scheme shown in Figure~\ref{fig:figure1}a, breaks the symmetry of the EIT interference and can produce a spectrally sharp bright resonance within the EIT lineshape. The ability to detect a lineshape change of the EIT transmission window is only limited by the laser linewidths, transit time broadening, shot noise and the decay and dephasing rates of the Rydberg states involved in the EIT process. The strength of the effect depends on the coupling to the RF electric field which is determined by the Rabi frequency of the transition, $\Omega_{RF} = \mu_{RF} E /\hbar$, where $E$ is the amplitude of the RF electric field and $\mu_{RF}$ is the transition dipole moment. Although technically challenging, it is possible, theoretically, to make RF electric field measurements at RF electric field strengths $\leq100\,$nV$\,$cm$^{-1}$. We know of no other atom based RF electric field sensor that relies on a quantum interferometric readout such as the one we have described here.

Rydberg atoms are highly excited atoms which are characterized by their principle quantum number $n$, and have long lifetimes \cite{Gallagher}. The transition dipole moment between neighboring Rydberg states scales as $n^2$ and can be extremely large, $\mu_{RF} \sim 4000\,ea_0$ for $n \sim 65$. The sensitivity to RF electric fields, then, is resonantly enhanced for frequencies in the range of 1-500$\,$GHz, as there are many resonant electric dipole transitions between Rydberg states in this frequency band. Because the RF electric field couples 2 close lying Rydberg states with a large transition dipole moment, $\Omega_{RF}$ can be large, $\sim 1\,$MHz, when the RF electric field is weak. As an example, consider the $^{87}$Rb $55D \rightarrow 54F$ transition at a frequency of $\sim 13.9\,$GHz. A beam intensity of $5\,$fW$\,$cm$^{-2}$ corresponding to an RF electric field amplitude of $64\,\mu$V$\,$cm$^{-1}$ yields $\Omega_{RF} \sim 1\,$MHz. A small RF electric field amplitude results in a signal that is straightforward to observe with current frequency stabilized diode lasers and room temperature vapor cells.

Because each atom is identical in structure our approach uses each noninteracting, independent atom that participates as an identical, stable quantum sensor of the electric field. The device is traceable because its sensitivity is directly linked to the properties of the atom, namely the atomic Rydberg wave functions or dipole moments which are well-known and can be determined even more precisely with modern spectroscopy, for example using frequency combs and ultracold atoms to better determine the Rydberg atom quantum defects. The properties of Rydberg atoms are readily calculated to high precision. Dipole moments can currently be determined to a level of $10^{-4}\,ea_0$ \cite{dipolemoments}, suggesting the Rydberg atomic dipole moments can be determined to $1$ part in $10^8$ using current methods.

\section{Discussion}

Measured probe laser transmission curves as a function of probe laser detuning for the 4-level system found in Figure~\ref{fig:figure1}a are shown in Figure~\ref{fig:figure2}. Traces for several different RF electric field amplitudes are displayed. At small RF electric field amplitudes, Figure~\ref{fig:figure2}a, the microwaves cause a broadening and reduction of the peak EIT transmission signal. The linewidth of the EIT transmission window is $\sim 4-5\,$MHz, while the relative drift of the coupling and probe lasers is $\leq 1\,$MHz determined from the measurements in the absence of an RF electric field. At intermediate RF electric field amplitudes, an absorption dip corresponding to the bright resonance appears within the EIT transmission window. The frequency for the transition is calculated to be $14.232\,$GHz using the quantum defects from \cite{Li03,Mack11}, in agreement with the experimentally observed frequency of $14.233\,$GHz. At large RF electric field amplitudes, Figure~\ref{fig:figure2}b,  the EIT transmission peak is split by the RF electric field. In this case, the peak splitting can be interpreted as Autler-Townes splitting of the $53D_{5/2}$ state and is relatively large because the calculated transition dipole moment for this transition is large, $\mu_\textup{RF}= 1.37 \times 10^{-26}\, \textup{C}\,\textup{m}$.  For comparison, $\mu_\textup{RF}$ is more than three orders of magnitude larger than the transition dipole moments on the coupling and probe transitions.  Without Doppler averaging the splitting of the peaks in Figure~\ref{fig:figure2}b is equal to the Autler-Townes splitting of the states, $\Omega_{RF}/2 \pi$. Doppler averaging changes the splitting of the peaks by the ratio of the probe laser to coupling laser wavelength. In our case, with $\lambda_\textup{p} \simeq 780\,$nm and $\lambda_\textup{c} \simeq 480\,$nm, the Doppler averaged peaks are separated by $1.625 \times \Omega_{RF} / 2 \pi$. In a vapor cell, the ratio of these wavelengths needs to be known accurately to make precise measurements of the RF electric field. The D2 transition wavelength is known to 1 part in $10^{11}$ \cite{Steck} and wavelengths can be determined to 1 part in $10^8$ with commercial laboratory wavemeters, so this effect is not limiting at this point. Using the Autler-Townes splitting and the calculated transition dipole moment, the electric field inside the cell is known with an uncertainty of $0.5\%$, determined, in our case, by the precision with which the splitting can be measured. The RF electric field strengths obtained in this way are in agreement with those calculated using the parameters of the resonant horn antenna including its position relative to the Rb vapor cell.  The difference between the two calculations is $10\%$. We believe the difference between the calculated and measured RF electric fields is mainly due to the uncertainty associated with the characterization of the horn antenna. Density matrix calculations for the 4-level system are also shown in Figure~\ref{fig:figure2}. The agreement between the 4-level theoretical calculations and the data show that the system is modeled well. Our full 52-level density matrix calculations which include all of the hyperfine states, $|F\,m_F\rangle$, of each of the participating levels, show that the populations of states $53D_{5/2}$ and $54P_{3/2}$ get optically pumped into the stretched $|4\,,4\rangle$ and $|3\,,3\rangle$ states, respectively, consistent with other works \cite{Bet11}. Our 4-level model uses these states for calculating the angular part of the  transition dipole moment. The attenuation and splitting of the EIT transmission peak shown in Figure~\ref{fig:figure2} is the signal we propose to use to precisely measure RF electric fields.

Figure~\ref{fig:figure3}a shows the peak height of the EIT transmission window as a function of RF electric field strength. In this experiment the lasers are locked to the EIT peak and the RF electric field strength is swept in time while the transmission of the probe beam is monitored. The red line shows the 4-level density matrix calculation for this range of RF electric field strengths. For these microwave fields the probe transmission increases from the 3-level EIT value.  A maximum increase of $\sim 4.5 \%$ is shown in Figure~\ref{fig:figure3}a. For our experimental parameters and RF electric field amplitudes less than 1400 $\mu$V $\textup{cm}^{-1}$, the probe transmission is increased in the presence of the microwaves with a maximum at $715\,\mu$V$\,$cm$^{-1}$. The increase in probe transmission is a result of the Doppler effect. When the RF electric field is small, the coupling to $54P_{3/2}$ is small and there is little population in this state. However, because the RF electric field wavelength is comparatively long and therefore less sensitive to Doppler shifts, the dominant effect of the microwave field is to shift, on average, the atoms participating in the EIT to smaller 2 photon detunings. This increases the EIT transmission signal on resonance. Figure~\ref{fig:figure3}b shows a plot of the calculated probe transmission as a function of velocity for $\Omega_{RF}=0\,$MHz and $\Omega_{RF}= 2 \pi \times 1\,$MHz to illustrate this effect. Although the probe transmission is suppressed for atoms near zero velocity, the suppression is more than made up for by the increased participation of atoms in the wings of the line. The position of the peak transmission in Figure~\ref{fig:figure3}a can be shifted by changing either the probe or coupling Rabi frequency.

Figure~\ref{fig:figure4} shows the central result of this paper. In Figure~\ref{fig:figure4}a, a series of traces of the change in probe laser transmission as a function of RF electric field frequency are presented for various RF electric field amplitudes. The microwave frequency is scanned 20 MHz in 200 ms to acquire these traces. The probe and coupling lasers are locked on the EIT transmission resonance. The increased transmission is due to the resonant microwave field and not an off-resonant AC Stark shift. When the microwave power is held constant and the RF electric field frequency is scanned over the $53D_{5/2} \rightarrow 54P_{3/2}$ resonance, the maximum signal occurs at the experimental transition frequency, $14.233\,$GHz. The smallest detected RF electric field shown in Figure~\ref{fig:figure4} is $8.33\,\mu$V$\,$cm$^{-1}$. Figure~\ref{fig:figure4}b shows the probe transmission peak heights as a function of RF electric field amplitude compiled for the data shown in Figure~\ref{fig:figure4}a. The solid red curve shows our 4-level density matrix calculation to be in excellent agreement with the data. For RF electric field amplitudes $< 100\,\mu$V$\,$cm$^{-1}$ the maximum probe transmission signal decays smoothly, as seen in Figure~\ref{fig:figure4}b.

To obtain an estimate of the sensitivity of our current apparatus, we measured the probe transmission on the EIT resonance and compared the signal in the cases where the RF electric field was on and then off. The coupling laser was modulated at $22\,$kHz while the RF electric field was modulated at $700\,$Hz. The signal was processed in 2 different lock-in amplifiers with an overall detection system integration time of $1\,$s. For a signal to noise ratio of $\sim 1$, we were able to measure an RF electric field amplitude of $\sim 20\, \mu$V$\,$cm$^{-1}$. The measurement corresponds to a sensitivity of $\sim 30\, \mu$V$\,$cm$^{-1} \, \sqrt{\mathrm{Hz}}^{-1}$.

\section{Conclusion}

In summary, we have demonstrated a new method for RF electrometry based on measuring a bright resonance in a thermal atomic vapor. This method is based on setting up each participating atom as a quantum interferometer. Using our approach, we were able to demonstrate a sensitivity to RF electric fields of $\sim 30\,\mu$V$\,$cm$^{-1} \sqrt{\mathrm{Hz}}^{-1}$ with a very modest setup, commonly found in modern atomic physics laboratories. We were able to detect RF electric fields as small as $\sim 8\,\mu$V$\,$cm$^{-1}$. The smallest detectable RF electric field amplitude demonstrated is below current antenna based standards used for RF electric field sensing and calibration. In addition, our method minimally perturbs the RF electric field, is a direct measure of the RF electric field linked to the Rydberg atom transition dipole moments, and is resonantly enhanced to give some immunity to out of band interference. There are many Rydberg atom transition frequencies in the range of $1-500\,$GHz that can be utilized for electric field amplitude measurements. The current apparatus can be improved in many ways, including the use of lasers with narrower spectral bandwidth, the use of frequency modulated spectroscopy, the implementation of lower noise photo-detectors, and the use of the dispersive nature of the sharp resonances used in our experiments. In addition, we speculate that if the technique is pushed to the quantum limit it may be possible to improve it by squeezing. We are currently focussing on using the refractive index changes associated with the narrow resonances observed in the 4-level system to improve our sensitivity. One can envision that these improvements, as well as a more detailed understanding of the 4-level Rydberg atom EIT system, can push the limits of this technique to $\lesssim 100\,$nV$\,$cm$^{-1}$, at least for some RF electric field frequencies. Now that quantum assisted sensing has been applied to RF electrometry, the outlook is bright for adopting such a method for future RF electric field standards. Such a standard hinges on knowledge of the transition dipole moments of the atoms used for the experiments which can currently be determined at the $10^{-4}\,ea_0$ level. The electric field can, in principle, then be measured to a precision of 1 part in $10^8$ using the Autler-Townes splitting of the Rydberg transitions, Figure~\ref{fig:figure2}b. In the future, Rydberg atom transition dipole moments and the other transition dipole moments involved in the process used here can be determined to much higher precision than currently known using cold atoms and frequency combs to measure quantum defects and make high resolution Stark shift measurements. These efforts may be catalyzed by efforts like ours to improve and adopt a quantum standard for RF electrometry.

\section{Methods}

\begin{figure}
    \includegraphics[width=\columnwidth]{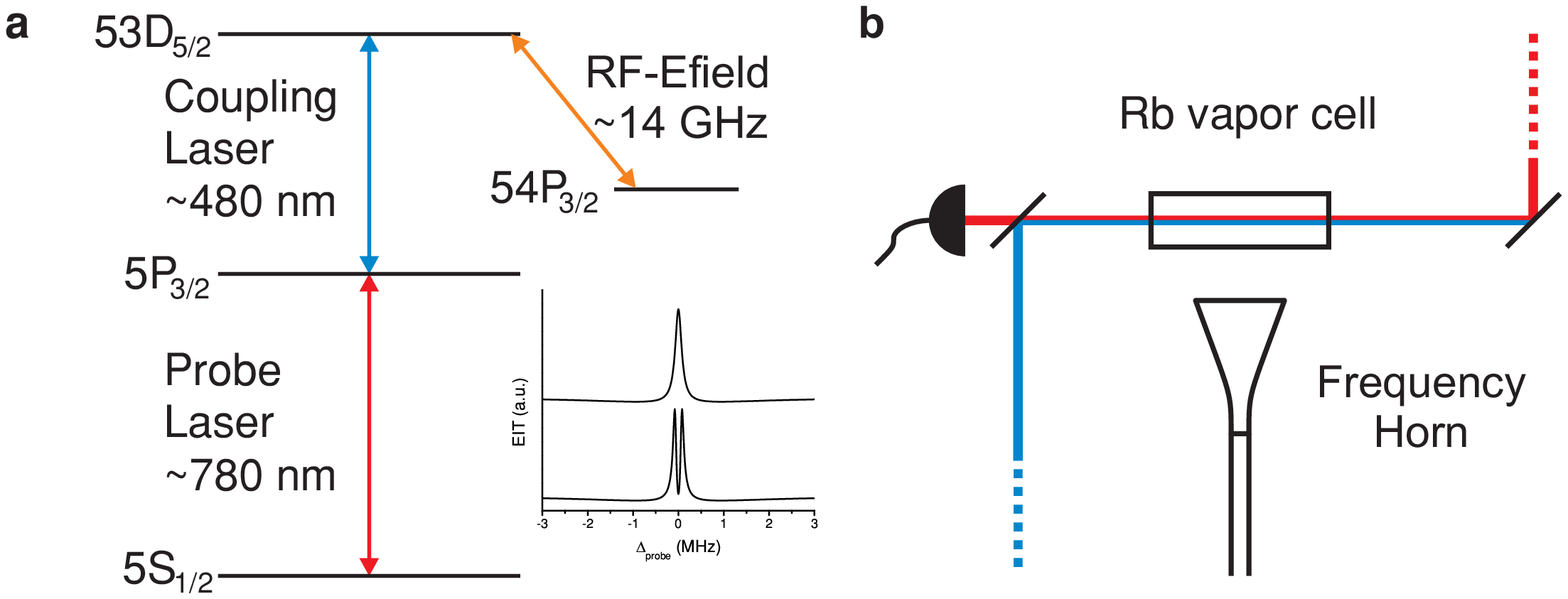}
   \caption{\textbf{Level diagram and experimental setup.} \textbf{a.)} The energy level diagram for the 4-level system used for the experiments in the paper. The top inset shows an example EIT feature associated the 3-level system without an RF electric field. The bottom inset shows an example of the bright resonance that is produced within the EIT window when an RF electric field is present. \textbf{b.)} The experimental setup used for the experiments.}
   \label{fig:figure1}
\end{figure}

\begin{figure}
    \includegraphics[width=\columnwidth]{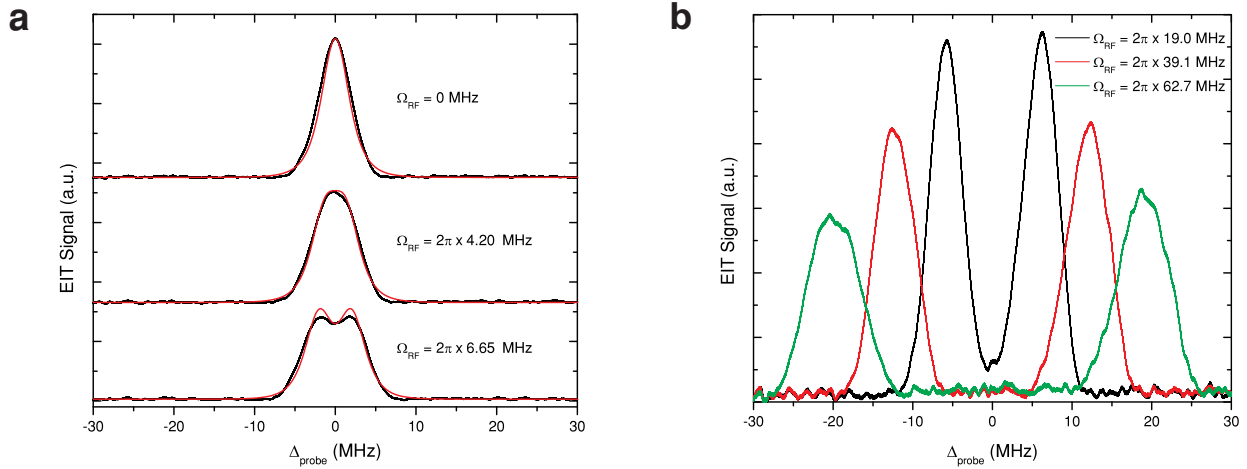}
   \caption{\textbf{Three level EIT and splitting from microwaves.} \textbf{a.)} The experimental bright resonance dip and attenuation of the EIT transmission signal for low RF electric field amplitudes (black) with theory curves (red).  \textbf{b.)}  Autler-Townes splitting of the $53D_{5/2} \rightarrow 54P_{3/2}$ Rydberg transition that occurs for larger RF electric field strengths. For all the graphs, the probe and coupling lasers are linearly polarized parallel to the RF electric field. The microwaves are linearly polarized perpendicular to the probe and coupling lasers propagation direction. The experimental parameters can be found in the methods section of the paper.}
   \label{fig:figure2}
\end{figure}

\begin{figure}
    \includegraphics[width=\columnwidth]{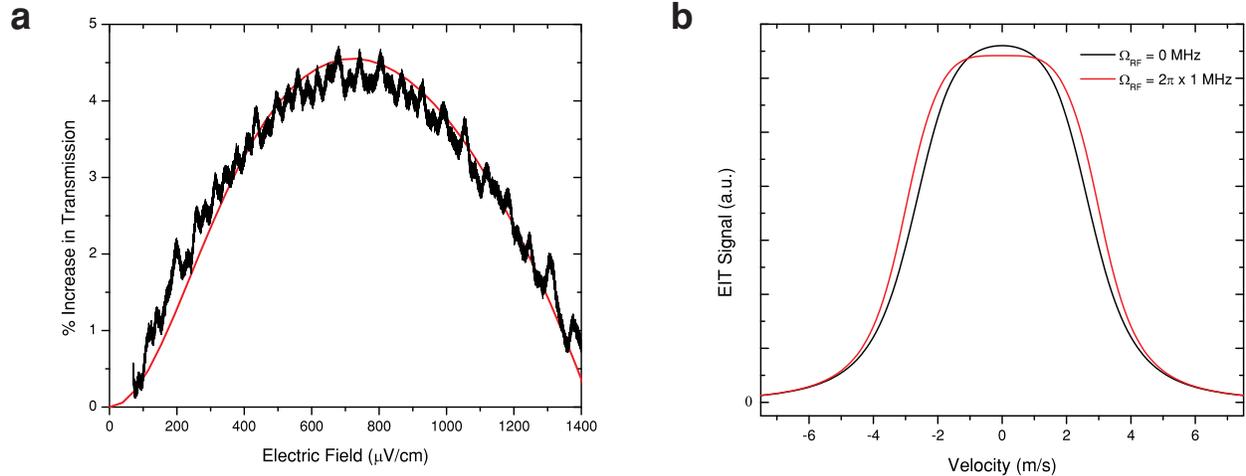}
   \caption{\textbf{4-level EIT signal vs. RF electric field amplitude.} \textbf{a.)} Transmission of the probe laser on resonance vs. RF electric field amplitude. The experimental results show agreement with a 4-level density matrix calculation. The red line is the theory and the black line is the experimental data. The parameters for the measurements are found in the methods section of the paper. The RF electric field is tuned $1\,$MHz off resonance. \textbf{b.)} This figure explains the enhancement of the EIT transmission when the RF electric field is applied for small RF electric field amplitudes. The RF electric field broadens the 2 photon EIT transition in velocity space in a thermal vapor. This leads to enhanced transmission of the probe laser. The probe and coupling lasers are circularly polarized with the same helicity relative to a quantization axis chosen to lie along the coupling or probe laser propagation direction.}
   \label{fig:figure3}
\end{figure}

\begin{figure}
        \includegraphics[width=\columnwidth]{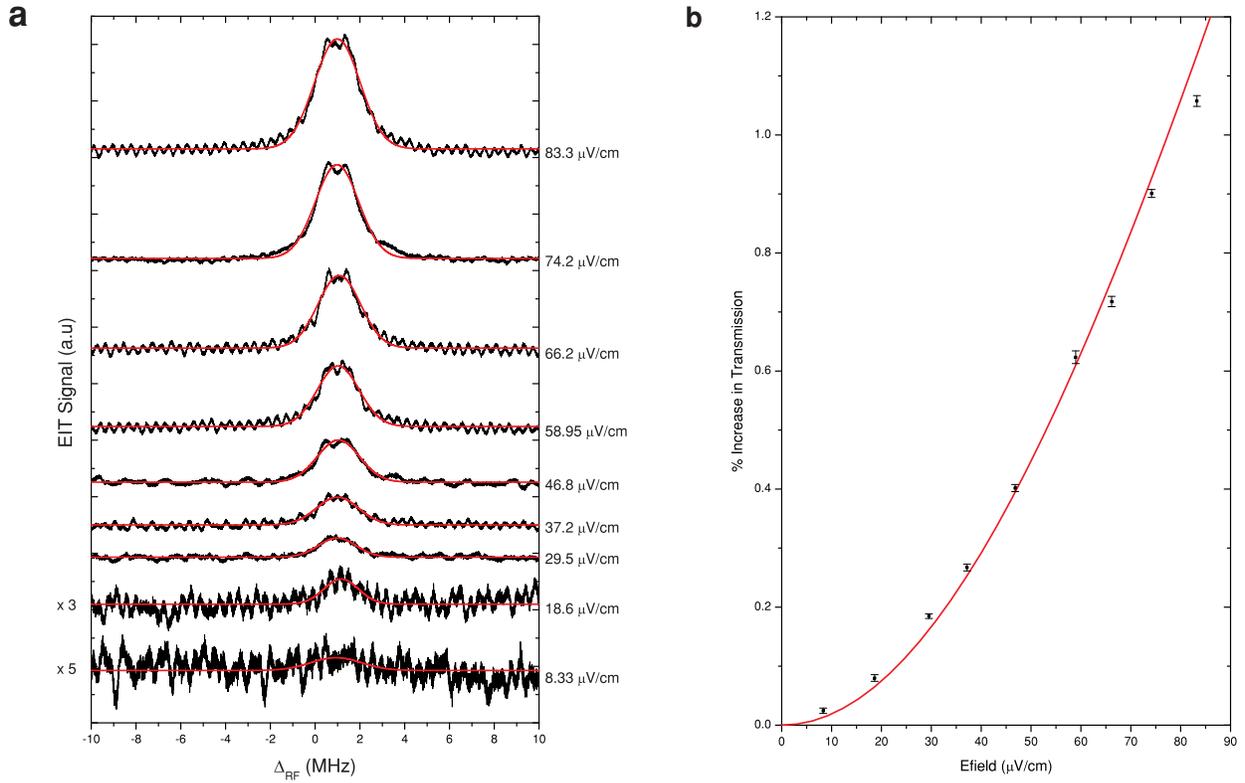}
    \caption{\textbf{4-level EIT vs. RF electric field amplitude.} \textbf{ a.)} Transmission lineshapes for different values of RF electric field amplitude as a function of RF electric field detuning.  The black curves are experimental data, and the red curves are Gaussian fits to the data. The graph shows the change in probe transmission as a function of RF electric field frequency.  All the data shown was taken under the same conditions as stated in Figure~\ref{fig:figure3}, except the RF electric field was tuned to resonance. Each trace is the average of 9000 scans. \textbf{b.)} This panel shows the on resonance peak height taken from the traces shown in panel a.) as a function of RF electric field amplitude.  The red line is a theoretical 4-level density matrix calculation. The error bars in the figure represent the standard deviation in the fits to the peak heights. The horizontal axis is derived from the RF power applied to the horn antenna and the associated experimental geometry.}
    \label{fig:figure4}
\end{figure}

The 4 level energy level diagram for the atomic system used in the experiments is shown in Figure~\ref{fig:figure1}a.  The probe and coupling lasers counter propagate through a room temperature $10\,$cm Rb vapor cell. The atom density is kept low enough so that collisions do not play a significant role in the experiments. In particular, we setup the experiment to avoid long range Rydberg atom interactions \cite{Schwettmann} and collisions involving Rydberg atoms. The average Rabi frequency on the probe transition is $\Omega_p = 2 \pi \times 6 \,$MHz and the average Rabi frequency on the coupling transition is $\Omega_c = 2 \pi \times 2\,$MHz. The coupling laser beam size is $100\,\mu$m and the probe beam size is $750\,\mu$m. Due to the focussing of the coupling laser, the effective interaction length is $7.5\,$cm. The probe laser, $\sim 780 \,$nm, is locked to the $5S_{1/2}(F=2) \rightarrow 5P_{3/2}(F=3)$ $^{87}$Rb transition. The coupling laser, $\sim 480\,$nm, is locked to a Fabry-Perot cavity. The Fabry-Perot cavity is stabilized to the EIT resonance using an independent Rb vapor cell. The $53D_{5/2} \rightarrow 54P_{3/2}$ Rydberg transition is driven with microwaves at $\sim 14.233\,$GHz. The microwaves are generated with an RF sweep generator (HP8340B). A resonant horn antenna is used to illuminate the experimental Rb vapor cell, perpendicular to propagation direction of the probe and coupling lasers, Figure~\ref{fig:figure1}b.  The probe and coupling beams are both circularly polarized with the same helicity relative to a quantization axis chosen to lie along the propagation direction of these 2 laser beams, except in Figure~\ref{fig:figure2} where they are both linear polarized parallel to the RF electric field. The microwaves are linearly polarized with the electric field oriented perpendicular to the propagation direction of the probe and coupling laser beams. The change in the probe laser transmission is detected using a standard silicon photodiode (Thorlabs DET10A), and processed using a lock-in amplifier. The intensity of the coupling beam is modulated at $22\,$kHz for the lock-in detection. A typical EIT trace is shown in Figure~\ref{fig:figure2}a for the 3-level Rydberg atom EIT without microwaves. We estimate the effective laser linewidths for the experiment to be $\sim 700\,$kHz from the convolution of the 2 photon linewidth and the laser lock error signals.  Figures~\ref{fig:figure2}a and \ref{fig:figure2}b show the effect of adding the RF electric field. The electric field measurements are carried out by detecting changes in the EIT transmission signal as a function of RF electric field strength. The data is obtained in 3 ways. First, the EIT signal at resonance, or maximum transmission, is monitored as a function of RF electric field strength, Figure~\ref{fig:figure3}. Second, the change in the EIT lineshape is recorded as a function of RF frequency for different RF electric field strengths, Figure~\ref{fig:figure4}. For the determination of the sensitivity, the probe transmission is monitored on resonance while modulating the RF electric field at a lower frequency. The signal is detected with a second lock-in to demodulate the RF electric field modulation with a total signal integration time of $1\,$s.

To model the system theoretically, we use a density matrix approach for the 4-level ladder system shown in Figure~\ref{fig:figure1}a \cite{fleisch05}.  The system is described by a master equation:
\begin{equation}\label{eq:1}
    \dot{\rho} = -\frac{i}{\hbar}[H,\rho] + L_d,
\end{equation}
where $H$ is the Hamiltonian for a 4-level ladder system in the interaction picture. $L_d$ is the standard dephasing and decay operator. For the calculations presented in this paper, $L_d$ includes spontaneous emission on the $5S_{1/2}(F=2) \rightarrow 5P_{3/2}(F=3)$ transition, $\Gamma_s = 2 \pi \times 6.1\,$MHz, and the effect of transit time broadening on the $53D_{5/2} \rightarrow 54P_{3/2}$ Rydberg transition, $\Gamma_{tt} = 2 \pi \times 270\,$kHz, \cite{thomas80}, because these terms dominate. Spontaneous emission, $\Gamma_R^{53D_{5/2}} \sim 2 \pi \times 1\,$kHz and $\Gamma_R^{54P_{3/2}} \sim 2 \pi \times 0.5\,$kHz,  black body decay, $\Gamma_{BB} \sim 2 \pi \times 1\,$kHz, and collisional dephasing rates, $\Gamma_{coll} \sim 2 \pi \times 1\,$kHz, of the Rydberg states are small compared to the transit time broadening for our experiment. The density matrix equations are solved in steady-state. These solutions are averaged over different velocities using a Maxwell-Boltzmann distribution at room temperature to account for the Doppler effect. Because the wavelength of the RF electric field is large, the Doppler effect on the $53D_{5/2} \rightarrow 54P_{3/2}$ transition can be neglected.

\begin{acknowledgments}
This work was supported by the DARPA Quasar program through a grant through ARO (60181-PH-DRP) and the NSF (PHY-1104424).
\end{acknowledgments}

\section{Author Contributions}
J.P.S. and T.P. conceived the idea. J.P.S. lead the project and wrote the paper. J.S., A.S., and H.K. carried out the experiments and reduced the data. A.S. and J.S. wrote the simulation programs. R.L. contributed useful ideas to the analysis of the experiment. All authors contributed extensively to the work.

\section{Competing Financial Interests}
The are no competing financial interests associated with this work.

\newpage

\end{document}